\documentclass[aps,pre,twocolumn,superscriptaddress, nobibnotes]{revtex4-2}
\usepackage{amsmath,amssymb}
\usepackage{graphicx}
\usepackage{hyperref}
\usepackage{physics}
\usepackage{bm}
\begin{document}

\title{Enhanced particle diffusion in fluctuating binary environments}

\author{Fivos Perakis}
\email{f.perakis@fysik.su.se}
\affiliation{Department of Physics, AlbaNova University Center, Stockholm University, S-106 91 Stockholm, Sweden}
\affiliation{Institute for Molecular Science, Myodaiji, Okazaki, Aichi 444-8585, Japan}

\author{Takeshi Kawasaki}
\affiliation{D3 Center, The University of Osaka, Toyonaka, Osaka, 560-0043, Japan}
\affiliation{Department of Physics, The University of Osaka, Toyonaka, Osaka, 560-0043, Japan}

\author{Shinji Saito}
\email{shinji@ims.ac.jp}
\affiliation{Institute for Molecular Science, Myodaiji, Okazaki, Aichi 444-8585, Japan}
\affiliation{The Graduate University for Advanced Studies (SOKENDAI), Myodaiji, Okazaki, Aichi 444-8585, Japan}

\begin{abstract}

We investigate single-particle diffusion in a two-state Langevin model where the friction coefficient randomly switches between low-friction (liquid-like) and high-friction (glassy-like) states. The dynamics are governed by the ratio between the friction switching time $\tau$ and the intrinsic velocity relaxation time $\tau_0$. For fast switching ($\tau/\tau_0 \lesssim 1$) the motion is homogeneous and Brownian, whereas for slow switching ($\tau/\tau_0 \gg 1$) the particle exhibits intermittent dynamics and an enhanced diffusion coefficient. Analysis of the single-particle overlap function $Q(t)$ and the dynamic susceptibility $\chi_4(t)$ reveals decoupling of the diffusion coefficient from the average friction upon cooling, which coincides with increasing temporal dynamic heterogeneity. This minimal model provides a transparent framework for understanding single-particle transport in media with fluctuating local mobility, including supercooled liquids and phase-separated soft materials.


\end{abstract}

\maketitle

Dynamical heterogeneity, where the local environment fluctuates both in space and time, is an essential feature of many complex physical and biological systems~\cite{ediger_spatially_2000}. Local fluctuations can fundamentally impact particle transport, often giving rise to dynamics that deviate significantly from classical Brownian motion in homogeneous media. Such behavior has been observed across a wide range of systems, including crowded biological environments~\cite{schurtenberger_observation_1989,fine_dynamic_1995, tanaka_universality_1996, tanaka_viscoelastic_2005, girelli_microscopic_2021} and the cytoplasm in cells~\cite{mcguffee_diffusion_2010, watson_macromolecular_2023}; glass forming liquids or  supercritical liquids near a critical point, with supercooled water as a prominent example~\cite{gallo_slow_1996, xu_relation_2005, kim_maxima_2017, saito_crucial_2018, debenedetti_second_2020}; colloidal suspensions in soft and active matter systems~\cite{ballesta_unexpected_2008, moller_velocity_2017}; magnetic domain fluctuations~\cite{seaberg_nanosecond_2017} and even diffusion in fluids under microgravity conditions~\cite{vailati_diffusion_2023}. Therefore, understanding the mechanisms driving such complex diffusion in heterogeneous and fluctuating landscapes is crucial for accurately describing transport in a broad spectrum of fluctuating environments.

Microscopic theories have long proposed that such anomalies arise from the coexistence of transient regions with distinct local mobilities, where particles experience intermittent transitions between slow and fast dynamical states\cite{stillinger_translation-rotation_1994, chang_heterogeneity_1997}. 
In this picture, transport is dominated by rare excursions through transiently fluid-like regions embedded in an otherwise viscous or solid-like matrix, producing large variations in individual particle displacements even when ensemble-averaged diffusion appears Brownian.
These microscopic fluctuations in the local environment, whether due to changes in viscosity, confinement, or interactions with dynamic surroundings, can dramatically influence particle trajectories~\cite{pacheco-pozo_langevin_2024, massignan_nonergodic_2014, van_den_broeck_noise-induced_1994, rozenfeld_brownian_1998, doering_resonant_1992}. As particles move through regions with distinct frictional properties, their mobility becomes intermittently enhanced or suppressed, leading to emergent behaviors that deviate from classical diffusion. 
This interplay between locally variable mobility and stochastic switching underlies the mechanism of "diffusing diffusivity"~\cite{jain_diffusing_2017, miyaguchi_generalized_2022}, offering a physically grounded framework for interpreting transport in systems exhibiting "Brownian yet non-Gaussian" characteristics~\cite{chubynsky_diffusing_2014}.

Here, we develop and analyze a simplified two-dimensional Langevin model with a two-state, time-dependent friction coefficient. 
Each particle switches stochastically between a low-friction ``liquid-like'' state with coefficient \(\gamma_l\) and a high-friction ``glass-like'' state with coefficient \(\gamma_g\).
The resulting dynamics are characterized through the mean-squared displacement (MSD) and diffusion coefficients, revealing that the mobility decouples for the local friction coefficient in the slow-switching regime. Moreover, analysis of the single-particle overlap function \(Q(t)\) and the associated four-point susceptibility \(\chi_4(t)\) directly connects this deviation to the emergence of heterogeneous dynamics on the single-particle level as the system approaches dynamical arrest.

We consider a single particle of mass $m$ evolving according to a two–state underdamped Langevin equation,
\begin{equation}
    m\,\dot{\mathbf{v}}(t) = -\gamma(t)\,\mathbf{v}(t) + \sqrt{2k_B T\,\gamma(t)}\,\boldsymbol{\xi}(t),
    \label{eq:langevin}
\end{equation}
where $\mathbf{r}(t)$ and $\mathbf{v}(t) = d\mathbf{r}/dt$ are the position and velocity vectors, $T$ is the temperature, and $k_B$ is the Boltzmann constant.  
The random force $\boldsymbol{\xi}(t)$ represents Gaussian white noise with zero mean and unit variance,
\begin{equation}
    \langle \xi_i(t)\rangle = 0, 
    \qquad 
    \langle \xi_i(t)\xi_j(t')\rangle = \delta_{ij}\,\delta(t-t').
    \label{eq:noise}
\end{equation}

The instantaneous friction coefficient $\gamma(t)$ takes one of two values,
\begin{equation}
    \gamma(t) =
    \begin{cases}
        \gamma_\mathrm{l}, & \text{liquid–like state},\\[4pt]
        \gamma_\mathrm{g}, & \text{glass–like state},
    \end{cases}
\end{equation}
and switches between them according to a Poisson process with mean waiting time $\tau$.  
The probability distribution of time intervals between switching events is therefore
\begin{equation}
    P(\Delta t_w) = \frac{1}{\tau}\exp\!\left[-\frac{\Delta t_w}{\tau}\right],
    \label{eq:poisson}
\end{equation}
which ensures continuous–time stochastic switching between the two frictional states. Here \(\Delta t_{\mathrm{w}}\) denotes the stochastic waiting time between successive state switches (Poisson process), which is distinct from the fixed integration step \(\Delta t\) used in the time-discretized dynamics below. 

Two characteristic time scales govern the dynamics:  
(i) the velocity relaxation time $\tau_0 = m/\langle\gamma\rangle$, which sets the decay of velocity correlations, and  
(ii) the mean switching time $\tau$, which characterizes the persistence of each frictional state.  
The ratio $\tau/\tau_0$ serves as a dimensionless control parameter.  
For $\tau/\tau_0 \lesssim 1$, the friction fluctuations are fast compared to the particle’s momentum relaxation, leading to effectively homogeneous Brownian motion.  
In contrast, for $\tau/\tau_0 \gg 1$, the friction varies slowly and the particle alternates between long‐lived high– and low–friction states, producing temporally heterogeneous dynamics and an enhancement of the long–time diffusion coefficient.

Eq.~(\ref{eq:langevin}) is integrated using a first–order (Euler–Maruyama)
discretization of the Langevin equation with piecewise–constant
$\gamma(t)$,
\begin{equation}
    \mathbf{v}_{n+1}
    = \mathbf{v}_n
      - \frac{\gamma_n}{m}\,\mathbf{v}_n\,\Delta t
      + \frac{\sqrt{2 k_B T \gamma_n \Delta t}}{m}\,\mathbf{R}_n,
\end{equation}
where $\mathbf{R}_n$ is a vector of independent, normally distributed
random numbers with zero mean and unit variance.
This explicit scheme is accurate provided that $\Delta t \ll \tau_0 = m/\langle\gamma\rangle$.
The particle positions are then updated as
\begin{equation}
    \mathbf{r}_{n+1} = \mathbf{r}_n + \mathbf{v}_{n+1}\,\Delta t,
    \label{eq:xupdate}
\end{equation}
which ensures accurate resolution of the velocity relaxation dynamics.

The switching dynamics of $\gamma(t)$ is implemented via pre–sampled
exponential waiting times drawn from Eq.~(\ref{eq:poisson}).
For each trajectory, a sequence of switching times $\{t_i\}$ is generated
up to the total simulation time $t_{\mathrm{max}}$, during which
$\gamma(t)$ alternates between $\gamma_\mathrm{l}$ and $\gamma_\mathrm{g}$ according
to the sampled intervals. Between successive switching events the friction remains constant.  
This approach provides an exact continuous–time representation of the stochastic switching process and avoids the discretization errors associated with fixed per–step switching probabilities. All simulations use reduced units (see Supplemental Material)  and a fixed integration time step $\Delta t = 10^{-3}$, ensuring $\Delta t \ll \min(\tau_0, \tau)$ for all parameter sets.  
Typical parameters are $\gamma_\mathrm{l}=1$, $\gamma_\mathrm{g}=100$, and switching times $\tau$ spanning several decades to explore both fast– and slow–switching regimes.
Initial states are randomly assigned between the liquid–like and glass–like friction values with an equal probability.

\begin{figure}[tbh]
    \centering
    \includegraphics[width=1\linewidth]{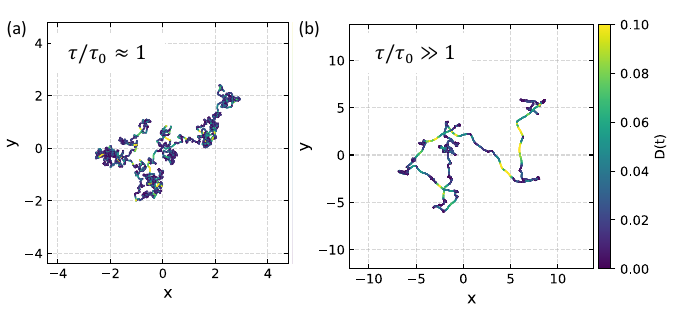}
    \caption{\label{fig:fig1}
Representative particle trajectories from simulations of the two–state Langevin model, color–coded by the instantaneous diffusion coefficient.  
(a) Fast–switching regime, $\tau/\tau_0 \approx 1$, corresponding to effectively homogeneous Brownian motion.  
(b) Slow–switching regime, $\tau/\tau_0 \gg 1$, where the particle exhibits extended displacements during long low–friction intervals, leading to temporally heterogeneous dynamics and enhanced diffusion.  
All trajectories are shown over the same total time $t_{\max}=100$ and simulation parameters $T=1$, $\gamma_\mathrm{l}=1$, and $\gamma_\mathrm{g}=100$.}
    \label{fig1}
\end{figure}

Figure~\ref{fig:fig1} shows representative trajectories obtained from the two-state Langevin simulations, where the instantaneous diffusion coefficient is encoded in the color scale. Simulations were performed for \( T=1 \). The panels are shown for a total trajectory time, \( t_\mathrm{max}=100 \). 

Figure~\ref{fig:fig1}~(a) corresponds to the fast–switching regime, $\tau/\tau_0 \lesssim 1$, in which friction fluctuations are on comparable timescales with the velocity relaxation.  
In this limit, the particle experiences an effectively averaged friction and undergoes homogeneous Brownian motion.  
Figure~\ref{fig:fig1}~(b) illustrates the slow–switching regime, $\tau/\tau_0 \gg 1$, where long residence times in the liquid–like (low–friction) state give rise to extended displacements and temporally heterogeneous dynamics.  
During these intervals, the particle exhibits correlated motion, resulting in enhanced long–time diffusion. These trajectories visually demonstrate how the interplay between the switching time $\tau$ and the velocity–relaxation time $\tau_0$ controls the crossover from homogeneous to heterogeneous dynamics in the two–state model.

\begin{figure}[tbh]
    \centering
    \includegraphics[width=1\linewidth]{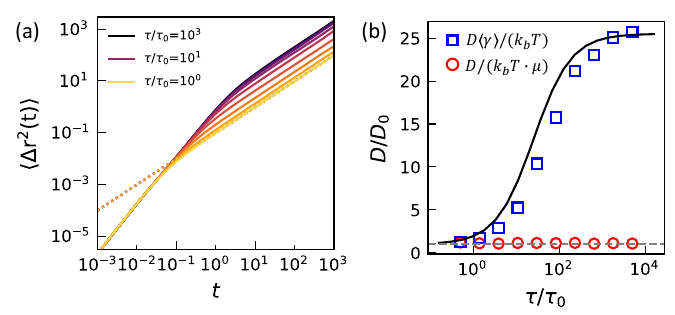}
    \caption{
    (a) Mean–squared displacement $\langle \Delta r^2(t) \rangle$ for different ratios $\tau/\tau_0$, illustrating the crossover from homogeneous to heterogeneous dynamics.  
    (b) Normalized transport coefficients: open circles show $D/(k_B T\,\mu)$ testing the Einstein relation, while open squares show $D\,\langle \gamma \rangle/(k_B T)$ testing the  relation $\mu = 1/\langle \gamma \rangle$. The solid line is calculated based on Eq.~\ref{eq:diffusion_enhancement}.  
    The Einstein relation holds for all $\tau/\tau_0$, whereas $\mu = 1/\langle \gamma \rangle$ fails in the slow–switching regime, leading to the observed enhancement of $D/D_0$.  
    }
    \label{fig:fig2}
\end{figure}

Figure~\ref{fig:fig2} summarizes the transport properties obtained from the two–state Langevin simulations at constant temperature $T=1$. In Fig.~\ref{fig:fig2}~(a) is shown the mean–squared displacement (MSD), $\langle \Delta r^2(t) \rangle$, as a function of time for different ratios $\tau/\tau_0$. 
Each curve represents an ensemble average over 100 independent trajectories of duration $t_{\max}=3\times10^3$. 
At short times, all curves exhibit a ballistic regime $\langle \Delta r^2(t) \rangle \propto t^2$, followed by a diffusive regime $\langle \Delta r^2(t) \rangle \propto t$ at long times. 
As $\tau/\tau_0$ increases, the crossover between the two regimes shifts to longer times, reflecting extended periods of inertial motion in the low–friction state. 

The apparent ballistic–like regime observed in the liquid–like state arises from the weak–damping limit ($\gamma_\mathrm{l} \ll \gamma_\mathrm{g}$) of the model, where the small friction coefficient leads to an extended velocity relaxation time $\tau_0 = m / \gamma_\mathrm{l}$.  
This regime should be interpreted as an idealized inertial limit of an underdamped tracer rather than literal ballistic motion of a finite–sized particle, since the model describes a point–like probe with an effective friction $\gamma(t)$. In this context, the observed ballistic–like behavior corresponds to a probe experiencing transiently low local friction. The long–time diffusion coefficient $D$ is obtained from linear fits to the MSD in the diffusive regime (dashed lines in Fig.~\ref{fig:fig2}a).

Figure~\ref{fig:fig2}~(b) shows the normalized transport coefficients extracted from these simulations.  
The open circles denote $D/(k_B T\,\mu)$, which tests the Einstein relation $D = k_B T\,\mu$ using the independently computed mobility $\mu$ obtained from the velocity autocorrelation function via the Green–Kubo relation (see Supplemental Material).  
The open squares show $D\,\langle \gamma \rangle / (k_B T)$, corresponding to the approximate relation $\mu = 1/\langle \gamma \rangle$.  
For fast switching ($\tau/\tau_0 \lesssim 1$), both ratios are close to unity, indicating that the Einstein relation holds and the dynamics are effectively homogeneous.  
In contrast, for slow switching ($\tau/\tau_0 \gg 1$), $D\,\langle \gamma \rangle / (k_B T)$ deviates strongly from unity while $D/(k_B T\,\mu)$ remains constant, demonstrating that the observed enhancement originates from the breakdown of $\mu = 1/\langle \gamma \rangle$ rather than from a violation of the Einstein’s law. The solid line shows the analytical prediction from Ref.~\cite{rozenfeld_brownian_1998}, based on the equation: 

\begin{equation}
\frac{D}{D_0} = 1 + \frac{\frac{1}{2} (\gamma_g - \gamma_l)^2}{2\, \gamma_g \gamma_l + \frac{m}{\tau}(\gamma_g + \gamma_l)},
\label{eq:diffusion_enhancement}
\end{equation}

\noindent
which quantitatively captures the observed behavior. We note that Eq.~\ref{eq:diffusion_enhancement} is reduced to simple Brownian motion for $\gamma_g = \gamma_l$. 

\begin{figure}[tbh]
    \centering
    \includegraphics[width=1.0\linewidth]{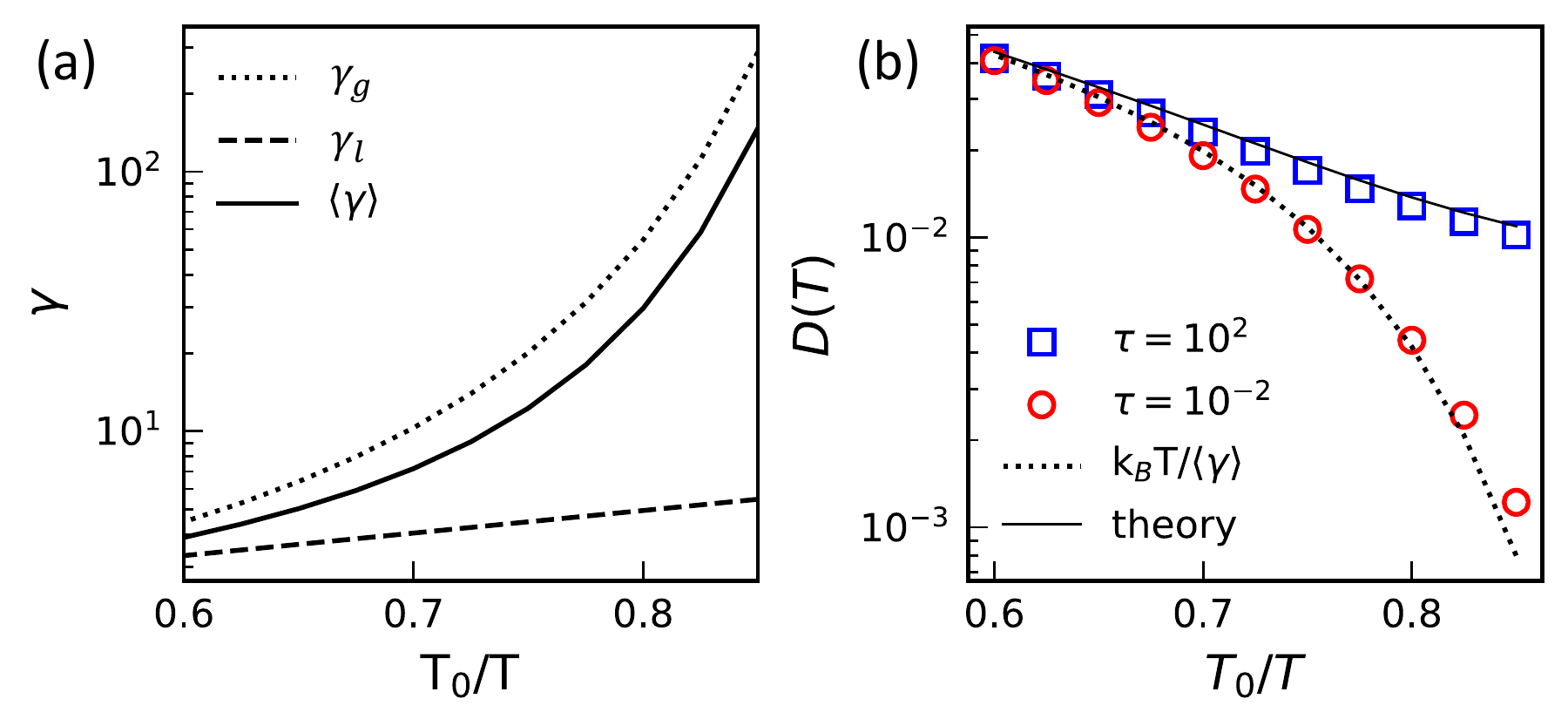}
    \caption{
    (a) Temperature dependence of the friction coefficients for the liquid–like (Arrhenius) and glass–like (VFT) states, together with their mean value $\langle\gamma(T)\rangle = \tfrac{1}{2}(\gamma_\mathrm{l}+\gamma_\mathrm{g})$.  
    (b) Diffusion coefficient $D$ as a function of reduced temperature $T_0/T$ for fast ($\tau/\tau_0 \lesssim 1$, red circles) and slow ($\tau/\tau_0 \gg 1$, blue squares) switching regimes.  
    The dashed line shows the relation $D_0 = k_B T / \langle\gamma(T)\rangle$ and the solid line is calculated based on Eq.~\ref{eq:diffusion_enhancement}. 
    Deviations in the slow–switching limit reflect the growing dominance of the high–friction state and the onset of heterogeneous dynamics.  
    }
    \label{fig:fig3}
\end{figure}

The temperature dependence of the two frictional environments is introduced through explicit functional forms for $\gamma_\mathrm{l}(T)$ and $\gamma_\mathrm{g}(T)$.  
The liquid–like friction follows an Arrhenius law,
\begin{equation}
\gamma_\mathrm{l}(T) = A_\mathrm{l}\exp\!\left(\frac{B_\mathrm{l}}{T}\right),
\end{equation}
characteristic of thermally activated transport in simple liquids,  
whereas the glass–like friction obeys a Vogel–Fulcher–Tammann (VFT) relation,
\begin{equation}
\gamma_\mathrm{g}(T) = A_\mathrm{g}\exp\!\left(\frac{B_\mathrm{g}}{T-T_0}\right),
\end{equation}
which diverges as $T \rightarrow T_0$ and captures the rapid dynamical slowdown near the glass transition.  
For the results presented here, we use $A_\mathrm{l}=1.0$ and $B_\mathrm{l}=0.2$ for the Arrhenius liquid, and $A_\mathrm{g}=1.0$, $B_\mathrm{g}=0.1$, and $T_0=0.1$ for the VFT glass.  
These parameters were selected such that $\gamma_\mathrm{l}(T)$ and $\gamma_\mathrm{g}(T)$ deviate significantly below $T_0/T \approx 0.5$, while preserving the qualitative behavior across a broad temperature range.

Figure~\ref{fig:fig3}(a) shows the temperature dependence of $\gamma_\mathrm{l}(T)$, $\gamma_\mathrm{g}(T)$, and their mean value $\langle\gamma\rangle = \tfrac{1}{2}(\gamma_\mathrm{l} + \gamma_\mathrm{g})$.  
The Arrhenius and VFT forms reproduce the expected trends of liquid–like and glass–like domains, respectively, with a pronounced divergence of $\gamma_\mathrm{g}(T)$ as $T$ approaches $T_0$. Figure~\ref{fig:fig3}(b) presents the diffusion coefficient $D$, obtained from the mean–squared displacement analysis, as a function of the reduced temperature $T_0/T$ for two representative regimes.  
The red circles correspond to the fast–switching regime ($\tau/\tau_0 \lesssim 1$) and the blue squares to the slow–switching regime ($\tau/\tau_0 \gg 1$).  
The dashed line indicates the relation $D_0 = k_B T / \langle\gamma\rangle$.  
In the fast–switching regime, the simulation data follow this relation closely, confirming that rapid environmental fluctuations average out the frictional heterogeneity.  
In contrast, for slow switching, $D$ deviates from $D_0$ around $T_0/T \approx 0.6$ and decreases less markedly as $T$ approaches $T_0$.  
This deviation reflects the increasing dominance of the low–friction (liquid-like) state which leads to the relative enhancement of the diffusion coefficient in comparison to the fast switching regime. 

\begin{figure}[tb]
    \centering
    \includegraphics[width=1.05\linewidth]{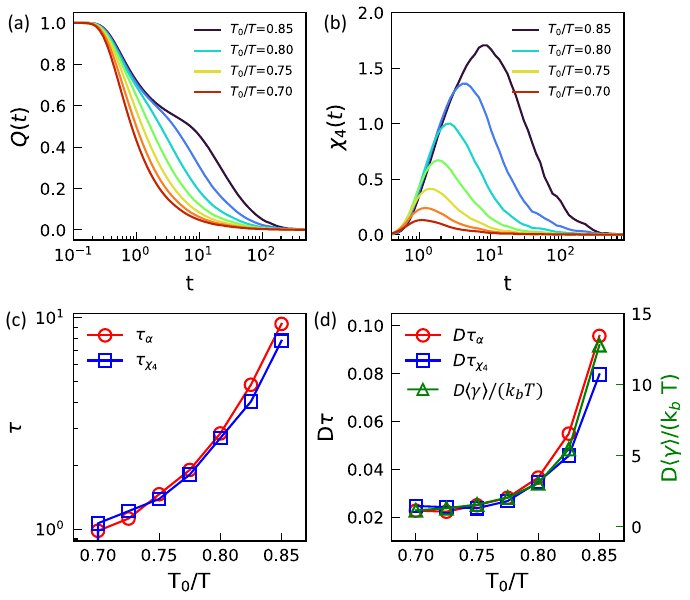}
    \caption{
    Analysis of dynamical heterogeneity at fixed switching time $\tau=100$. 
    (a) Single-particle overlap function $Q(t)$ for several temperatures, showing slower decay and a pronounced plateau upon cooling. 
    (b) Four-point susceptibility $\chi_4(t)$ with an increasing peak height and shift to longer times at lower $T$, reflecting enhanced temporal intermittency. 
    (c) Comparison of relaxation times $\tau_\alpha$ (red circles) from $Q(t)$ and $\tau_{\chi_4}$ (blue squares) from the $\chi_4(t)$ peak, showing similar temperature dependence. 
    (d) The $\mu = 1/\langle \gamma \rangle$ deviation quantified via $D\tau_\alpha$ (red circles), $D\tau_{\chi_4}$ (blue squares), and $D\langle\gamma\rangle/T$ (green triangles, right axis). 
    All three quantities increase below $T_0/T \approx 0.7$, indicating enhanced diffusion and heterogeneous dynamics in the slow-switching regime.
    }
    \label{fig:fig4}
\end{figure}

In Fig.~\ref{fig:fig4} we analyze dynamical heterogeneity in the slow-switching regime by fixing the switching time to $\tau = 100$ and varying the temperature $T$. 
The single-particle overlap function \(Q(t)\) and the corresponding four-point susceptibility \(\chi_4(t)\) provide complementary information on relaxation times and temporal fluctuations of the dynamics~\cite{lacevic_spatially_2003}. 
The overlap function \(Q(t)\) measures the fraction of trajectories that remain within a cutoff distance \(a\) of their initial position after a time \(t\). 
For each trajectory,
\begin{equation}
Q_i(t) = \big\langle w\big(|\mathbf{r}_i(t_0+t) - \mathbf{r}_i(t_0)|\big) \big\rangle_{t_0},
\end{equation}
where $w(r)=1$ if $r<a$ and $w(r)=0$ otherwise, and $\langle\cdots\rangle_{t_0}$ denotes averaging over time origins $t_0$. 
The ensemble-averaged overlap is
\begin{equation}
Q(t) = \frac{1}{N}\sum_{i=1}^N Q_i(t),
\end{equation}
with $N$ the number of independent trajectories.  
Unless otherwise noted, the cutoff distance is $a=0.2$ (see Supplemental Material).  
Figure~\ref{fig:fig4}(a) shows that upon cooling, $Q(t)$ decays more slowly and develops a pronounced plateau, indicating transient localization in the high-friction (glassy-like) state. The overlap function $Q(t)$ is the real--space analogue of the self--intermediate scattering function $F(k,t)$, shown in the Supplemental Material, confirming that both observables yield identical relaxation behavior.

The associated four-point susceptibility quantifies fluctuations of the overlap function and is defined as
\begin{equation}
\chi_4(t) = N \left[ \langle Q_i(t)^2\rangle - \langle Q_i(t)\rangle^2 \right].
\end{equation}
In the present single-particle model, $\chi_4(t)$ does not measure spatial correlations but rather captures the \emph{temporal intermittency} of individual trajectories, i.e. the fluctuations between slow and fast dynamical episodes induced by time-dependent friction.
As shown in Fig.~\ref{fig:fig4}(b), the $\chi_4(t)$ peak height increases and shifts to longer times as $T$ decreases, reflecting increasingly intermittent dynamics at low temperature.

Figure~\ref{fig:fig4}(c) compares two characteristic timescales: the relaxation time $\tau_\alpha$, defined by $Q(\tau_\alpha)=0.45$ (red circles), and the time $\tau_{\chi_4}$ corresponding to the $\chi_4(t)$ peak (blue squares). 
Both quantities exhibit similar temperature dependence, demonstrating that the slow relaxation and temporal heterogeneity occur on comparable timescales. 
Finally, Fig.~\ref{fig:fig4}(d) illustrates the origin of the $\mu = 1/\langle \gamma \rangle$ deviation~\cite{kawasaki_identifying_2017} by plotting $D\tau_\alpha$ (red circles), $D\tau_{\chi_4}$ (blue squares), and $D\langle\gamma\rangle/T$ (green triangles, right axis) as functions of $T_0/T$. 
All three quantities increase sharply below $T_0/T \approx 0.7$, signaling enhanced diffusion relative to the time-averaged friction $\langle\gamma\rangle$ as the system approaches dynamical arrest.
In contrast, for fast switching (see Supplemental Material), $D\langle\gamma\rangle/T$ remains essentially constant, consistent with homogeneous Brownian motion.

The results of this work establish a minimal yet general framework for understanding deviations of the $\mu = 1/\langle \gamma \rangle$ relation in systems with temporally fluctuating friction. 
In the classical picture, the $\mu = 1/\langle \gamma \rangle$ relation predicts that the diffusion coefficient $D$ scales inversely with $D = k_B T / \langle \gamma \rangle$ when the friction $\gamma$ is constant. 
In the present two-state Langevin model, $\gamma(t)$ switches stochastically between low- and high-friction values, representing intermittent access to fast (liquid-like) and slow (glassy-like) environments. 
Our simulations show that when the friction switching time $\tau$ becomes comparable to or larger than the intrinsic velocity relaxation time $\tau_0 = m / \langle \gamma \rangle$, the long-time diffusion coefficient $D$ becomes decoupled from the time-averaged friction $\langle \gamma \rangle$. 
Instead, the system obeys the Einstein relation $D = k_B T\,\mu$, where the mobility $\mu$ is enhanced relative to $1/\langle \gamma \rangle$ because the particle spends extended periods in the low-friction state. 

The model therefore provides a generic description of transport in heterogeneous or fluctuating environments. 
Although formulated as a single-particle process, it captures the essential physics of intermittently mobile dynamics observed in a variety of complex fluids.
Examples include tracer diffusion in supercooled or critical liquids with coexisting slow and fast domains, such as the interconversion between low- and high-density liquid states in supercooled water~\cite{xu_relation_2005, kim_maxima_2017, saito_crucial_2018, pathak_temperature_2019, kim_experimental_2020, debenedetti_second_2020}, or the dynamics of solutes and macromolecules in viscoelastic and phase-separating soft materials~\cite{schurtenberger_observation_1989,fine_dynamic_1995, tanaka_universality_1996, tanaka_viscoelastic_2005, girelli_microscopic_2021, hong_behavior_2020}. 
In such systems, local variations in viscosity or density give rise to time-dependent frictional environments analogous to those modeled here.

By introducing the ratio $\tau/\tau_0$ as a dimensionless control parameter, our simulations reveal a clear dynamical crossover: for $\tau/\tau_0 \lesssim 1$, the friction fluctuates rapidly and the dynamics are homogeneous and Brownian, while for $\tau/\tau_0 \gg 1$, the particle exhibits intermittent, heterogeneous motion with an enhanced effective diffusion coefficient. Analysis of the single-particle overlap function $Q(t)$ and the associated susceptibility $\chi_4(t)$ demonstrates that the enhancement of diffusion at low temperatures is accompanied by increasingly intermittent trajectories and stretched relaxation. 
This temporal heterogeneity mirrors the phenomenology of dynamic heterogeneity in glass-forming liquids, even though spatial correlations are absent. 
Hence, the model isolates the essential mechanism by which slow environmental fluctuations and heterogeneous frictional landscapes produce decoupling of mobility $\mu$ from the average friction $\langle \gamma \rangle$.

\section*{Acknowledgments}
We would like to acknowledge Zhiye Tang for useful discussion and comments. F.P. acknowledges financial support by the Swedish National Research Council (Vetenskapsrådet) under Grant No. 2019-05542, 2023-05339 and by Knut och Alice Wallenberg foundation (WAF, Grant. No. 2023.0052). S.S. acknowledges financial support by JSPS through the Grant-in-Aid for Scientific Research (JP21H04676 and JP23K17361). T.K. acknowledges support by the JST FOREST Program (Grant No. JPMJFR212T), AMED Moonshot Program (Grant No. JP22zf0127009), JSPS KAKENHI (Grant No. JP24H02203), and Takeda Science Foundation.

\bibliography{ref_abbrev}

\end{document}